\documentclass[twocolumn,showpacs,floatfix,prb,footinbib]{revtex4}

\usepackage{amssymb}
\usepackage{graphicx}
\usepackage{dcolumn}
\usepackage{bm}
\begin{document}

\title{Magnetoconductance through a vibrating molecule in the Kondo regime}
\author{P. S. Cornaglia}
\author{D. R. Grempel}
\affiliation{CEA-Saclay, DSM/DRECAM/SPCSI, B\^at. 462, F-91191
Gif-sur-Yvette, France}

\begin{abstract}
The effect of a magnetic field on the equilibrium spectral and
transport properties of a single-molecule junction is studied
using the numerical renormalization group method. The molecule is
described by the Anderson-Holstein model in which a single
vibrational mode is coupled to the electron density. The effect of
an applied magnetic field on the conductance in the Kondo regime
is qualitatively different in the weak and strong electron-phonon
coupling regimes. In the former case, the Kondo resonance is split
and the conductance is strongly suppressed by a magnetic field $g
\mu_B B\gtrsim k_BT_{\rm K}$, with $T_{\rm K}$ the Kondo
temperature. In the strong electron-phonon coupling regime a
charge analog of the Kondo effect develops. In this case the Kondo
resonance is not split by the field and the conductance in the
Kondo regime is enhanced in a broad range of values of $B$.
\end{abstract}


\maketitle
\section{Introduction}
Transport through molecular nanodevices\cite{Nitzan2003} has recently been a
subject of intense experimental
 \cite{Reed1997,HPark_2000,JPark_2002,WLiang_2002,Kubatkin_2003,Yu2004,Pasupathy2004,Yu2004b}
 and theoretical
 \cite{Ness2001,Zhang2002,Braig_2003,Flensberg_2003,Mitra_2003,Cornaglia2004,Cornaglia2005a,Chen2004}
 investigation. These systems are not only of
great interest because of their potential for technological
applications but they also pose some interesting an challenging
theoretical problems. They arise because in molecular nanodevices
electron-electron ({\it e--e}) and electron-phonon ({\it e--ph})
interactions are present simultaneously and their interplay may
lead to new physics if their strengths are comparable.

Effects of {\it e--e} interactions in single-molecule devices were
seen through the observation of Coulomb blockade peaks in the
conductance. At these peaks the charge of the molecule changes by
one electron while it stays well defined in the valleys between
them. \cite{Kubatkin_2003}

The Kondo effect \cite{Hewson-book} was also observed in C$_{60}$
single molecules coupled to metallic \cite{Yu2004} and
ferromagnetic electrodes.\cite{Pasupathy2004} It occurs in the
valleys in which the charge of the molecule is close to an odd
integer. Its experimental manifestation is an increase of the
conductance with decreasing temperature and a zero-bias peak in
the differential conductance below a characteristic temperature
scale, the Kondo temperature $T_{\rm K}$.

Furthermore, some features in the I-V characteristics of a
vibrating C$_{60}$ single-molecule transistor were interpreted as
being a manifestation of the effects of the {\it e--ph}
interaction. \cite{HPark_2000}

It was shown in previous work~\cite{Cornaglia2004,Cornaglia2005a}
that the interplay of {\it e--e} and {\it e--ph} interactions has
important effects on the physical properties of the system in the
Kondo regime. There are two different cases characterized by the
strength of the {\it e--ph} coupling. When the latter is weak (in
a sense that will be defined below), the ground state of the
isolated molecule in the odd electron number valleys is a spin
doublet. Then, the conventional Kondo effect appears when the
molecule is coupled to metallic leads. The presence of the {\it
e--ph} interaction results in an increase of the spin fluctuation
energy and of the amplitude of charge fluctuations. In the strong
{\it e--ph} coupling regime, polaronic effects result in an
effective attractive interaction $U_{\rm eff}$ between the
electrons and this leads to qualitatively new physics. If
the charge in the device is close to an odd integer the ground
state of the isolated molecule is a spinless charge doublet. This
is the regime in which the charge Kondo effect develops. It
differs from the conventional Kondo effect in that the low lying
excitations of the system are charge fluctuations whereas spin
fluctuations become increasingly stiff as the {\it e--ph} coupling
increases. The charge Kondo effect enhances the transport through
the molecule and leads to the appearance of a narrow peak in the
conductance as a function of the gate voltage.

In this paper we study the properties of the device in an applied
magnetic field. We found that these are quite different at weak
and strong {\it e--ph} coupling. At weak coupling the system
behaves qualitatively as in the absence of the latter.
\cite{Costi2001} The magnetic field breaks the spin symmetry of
the molecule's ground state and splits the Kondo peak in the
density of states thus suppressing the conductance for fields
$g\mu_BB\gtrsim k_B T_{\rm K}$. At strong coupling, the molecule's
low lying states are spinless and are thus insensitive to the
magnetic field. The Kondo peak is not split by the latter and
there is almost perfect transmission at low temperatures for
fields $B\lesssim B_{\rm crit} = \left|U_{\rm eff}\right|/g\mu_B$.
Furthermore, within this range of field values, the width of the
conductance vs. gate voltage peak increases rapidly with $B$.
Beyond $B_{\rm crit}$ there is level crossing and the conductance
peak splits into two peaks that are strongly spin polarized.

The rest of this paper is organized as follows. In Section
\ref{sec:model} we present the model Hamiltonian and some general
considerations that follow from exact Fermi liquid relationships
at $T=0$. In Section \ref{sec:numerical} we present Numerical
Renormalization Group results for the spin resolved spectral
densities and conductance in the presence of a local magnetic
field at finite temperature and as a function of gate voltage. A
summary and our concluding remarks are presented in Section
\ref{sec:conclusions}.

\section{Model Hamiltonian}\label{sec:model}
We consider a model of a molecule with a single relevant
electronic level coupled to left (L) and right (R) metallic
electrodes and a vibrational mode of frequency $\omega_0$ linearly
coupled to the charge fluctuations of the molecule. The
Hamiltonian of the system is
\begin{equation}
H=H_{M} + H_{E} + H_{M-E}+H_{MF},
\label{hamil}
\end{equation}
where the first three terms describe the isolated molecule, the
electrodes and  their coupling, respectively.  The last term is
the coupling of the spin of the molecule to an external magnetic
field acting only on the molecule. We have
\begin{eqnarray}
H_{M}&=&\varepsilon_d \hat{n}_d + U \hat{n}_{d \uparrow}
\hat{n}_{d \downarrow}-\lambda\;
\left(\hat{n}_d-1 \right)\left(a^{}+a^{\dagger}\right)\nonumber\\
&+&\omega_0a^{\dagger}a^{},\\
H_{E}&=&\sum_{{\bf k},\sigma,\alpha=\{{\rm L,R}\}}
\varepsilon_{\sigma\alpha}({\bf k})\;
c^{\dagger}_{{\bf k} \sigma \alpha} c^{}_{{\bf k} \sigma
\alpha},\\
H_{M-E}&=&\sum_{{\bf k},\sigma,\alpha=\{{\rm L,R}\}}\;V_{{\bf
k}\sigma\alpha}\left(d^{\dagger}_{\sigma}
\;c^{}_{{\bf k}\sigma\alpha} + c^{\dagger}_{{\bf
k}\sigma\alpha}\;d^{}_\sigma \right),\\
H_{MF}&=& g\mu_B B \hat{S}_z^d.
\end{eqnarray}
Here, $\hat{n}_d = \sum_{\sigma} d^{\dagger}_\sigma d^{}_\sigma$
and $\hat{S}_z^d=\frac{1}{2}(d_{\uparrow}^\dagger d_{\uparrow} -
d_{\downarrow}^\dagger d_ {\downarrow})$ are the charge and spin
operators of the molecule, respectively.  The position of the
electronic molecular level relative to the Fermi level of the
electrodes is denoted by $\varepsilon_d$ and $U$ is the Coulomb
repulsion between two electrons that occupy the same molecular
level. Finally, $a^\dag$ is the creation operator of a phonon
of frequency $\omega_0$ and
$\lambda$ is the {\it {\it e--ph}} coupling constant.
We consider for simplicity the case of identical electrodes
with dispersion $\varepsilon_{\rm{R}}({\bf k})
=\varepsilon_{\rm{L}}({\bf k}) =\varepsilon({\bf k})$ and a
${\bf k}$-independent tunnelling matrix element.  We further
assume that the conduction band is symmetric around the chemical
potential and denote its half-width by $D$. In the
following we use units such that $k_B=D=g\mu_B=1$.

The eigenstates of the Hamiltonian $H_M+H_{MF}$ of the isolated
molecule are direct products of electronic states (denoted by a
subscript $e$) and oscillator states. The eigenfunctions  and
their energies can be written explicitly:
\begin{eqnarray}\label{eq:enes}
\begin{array}{ll}
\left |0,m\right> = \tilde{U}^{-}\left |0\right>_e
\left|m\right>\!, & E^0_m=-\frac{\lambda^2}{ \omega_0} +
 m\omega_0 ,\\
\left |\sigma,m\right> = \left |\sigma\right>_e\left|m\right>\!,
 & E^\sigma_m=\varepsilon_d + \sigma B
+ m\omega_0,\\
\left |2,m\right> = \tilde{U}^{+}\left |\uparrow
\downarrow\right>_e\left|m\right>\!, & E^2_m=-{\lambda
^2\over \omega_0} + 2\varepsilon_d + U + m\omega_0,
\end{array}
\end{eqnarray}
where $\tilde{U}^{\pm} =\exp{\left[\pm{\lambda/\omega_0}
\left(a^{\dagger}-a\right)\right]}$, $\left|m\right>$ is the
$m$-th excited state of the harmonic oscillator and $\sigma = \pm
1/2$. The ground state energies of the empty and doubly occupied
electronic states are reduced by the polaronic shift
$\lambda^2/\omega_0$.

At zero field the differences between the occupied and empty
states ${\tilde{E}^\sigma_m} = E^\sigma_m - E^0_m$ and
${\tilde{E}^2_m}- E^0_m$ are ${\tilde{E}^\sigma_m}= \varepsilon_d
+ \lambda^2/\omega_0$ and ${\tilde{E}^2_m}= 2\varepsilon_d + U$,
respectively. The effective interaction between two electrons
occupying the same state is thus ${\tilde{E}^2_m} - 2
{\tilde{E}^\sigma_m} = U- 2\lambda^2/\omega_0 \equiv
U_{\rm eff}$.

The sign of this quantity determines the low energy physics of the
system and its response to an applied field. For positive
$U_{\rm eff}$, the ground state of the isolated molecule is a
spin doublet and the Kondo effect occurs at low temperatures when
the molecule is coupled to the leads.
For negative $U_{\rm eff}$ the ground state of the isolated
molecule is either empty, doubly occupied or a charge doublet. In
the latter case the coupling of the molecule to the leads
generates the charge Kondo effect with an anisotropic Kondo (AK)
exchange coupling.
\cite{Schuttler1988,Cornaglia2004,Cornaglia2005a}

 In this paper we only study the case of
paramagnetic contacts and, therefore, we set $\Gamma_{\sigma
\textsc l}(\omega)=\Gamma_{\sigma \textsc
r}(\omega)\equiv \frac{1}{2}\Gamma(\omega)$, where $\Gamma(\omega)\equiv 2\pi \sum_k
\delta\left(\omega-\varepsilon({\bf k})\right)V_{{\bf k}}^2$.
The spin resolved conductance of the
molecular junction at zero bias
is~\cite{Hershfield1991,Pastawsky1991,Pastawsky1992,Meir1992,Jauho1994}
\begin{equation}
\label{conductance}
G_\sigma={e^2\over h}\ \pi
 \int_{-\infty}^{\infty} d\omega
 \left(-{\partial f(\omega)\over \partial \omega}\right)
 \rho_{d\sigma}({\omega})\Gamma(\omega),
\end{equation}
where $\rho_{d\sigma}(\omega)=-\pi^{-1}{\rm Im}{\mathcal
G}_{d\sigma}(\omega)$, $ {\mathcal
 G}_{d\sigma}(\omega)$ is the retarded electronic Green function of
the molecule in the presence of the leads and $f(\omega)$ is the
Fermi distribution.

At zero temperature, in the wideband limit, the dimensionless
conductance satisfies the Fermi liquid relationship
\begin{equation}
\label{conductance-T0} g\equiv {G \over G_0}=\pi \sum_\sigma
\Gamma\;\rho_{d\sigma}(0)= \sum_\sigma\sin^2\left(\pi
n_{d\sigma}\right),
\end{equation}
where $G_0=e^2/h$ is the quantum of conductance, $n_{d\sigma} =
\langle \hat{n}_{d \sigma}\rangle$, $\Gamma=\Gamma(0)$, and the second equality follows
from Luttinger's theorem \cite{Hewson-book} generalized to the
case in which {\it e--ph} interactions are present.
\cite{Cornaglia2005a}

Consider first the case $\sum_\sigma n_{d\sigma}=1$
corresponding to the electron-hole symmetric situation
$\varepsilon_d = -U/2$. The conductance in zero field is then
$g=2$. In the presence of the applied field $B$ the molecule
acquires a magnetization $m_d=\frac{1}{2}
\left(n_{d\downarrow}-n_{d\uparrow}\right)$ and
Eq.~(\ref{conductance-T0}) becomes
\begin{equation}
 g= \sum_\sigma\sin^2[\pi (1/2 - 2\sigma  m_d)] = 2 \cos^2\left(\pi m_d\right).
\label{eq:luttinger-m}
\end{equation}
In the low field limit we have,
\begin{equation}
 g \approx 2  (1- \pi^2 m_d^2)\approx  2 (1- \pi^2 \chi_d^2  B^2),
\label{g-de-B}
\end{equation}
were $\chi_d=\partial m_d/\partial B$ is the local spin
susceptibility. In the large field limit we have $m_d \to 1/2$ and the
conductance vanishes. The characteristic field beyond which the
conductance is suppressed is $B^{\star} \sim \chi^{-1}_d$.

For $U_{\rm eff}$ large and positive $\chi^{-1}_d \sim T_{\rm K}$
and increases slowly with increasing $\lambda$.
\cite{Cornaglia2004,Cornaglia2005a} Therefore, $B^{\star} \sim
T_{\rm K}$ is small in the limit $U_{\rm eff} \gg \Gamma$. For $B
\gtrsim B^{\star}$ the conductance at $\varepsilon_d\sim -U/2$ is
thus suppressed and $g$ acquires a double-peak structure as a
function of the gate voltage.
 The zero temperature conductance at the positions of the peaks is $g=1$.
\cite{Costi2001}

For $U_{\rm eff}$ large and negative $\chi^{-1}_d \sim
\left|U_{\rm eff}\right|$  and
$B^{\star}$ is thus large.\cite{Cornaglia2004,Cornaglia2005a} In this case, a single conductance peak
with $g=2$ is expected in a broad range of magnetic fields.

In zero field but off electron-hole symmetry, $\varepsilon_d=-U/2
+ \delta \varepsilon_d$, we have:
\begin{equation}
g \approx 2 \;(1- \pi^2 \delta n_d^2/4)\approx 2 \;(1- \pi^2
\chi_c^2 \delta\varepsilon_d^2/4), \label{g-de--e}
\end{equation}
where $\chi_c=-\partial n /\partial \varepsilon_d$ is the charge
susceptibility and $\delta n_d = \sum_\sigma n_{d\sigma} - 1$. The
width $\delta \varepsilon^{\star}_d$ of the conductance peak is
$\sim\!\chi^{-1}_c\!\sim\!T_{{\rm {AK}}}$, the Kondo temperature
of the effective anisotropic Kondo model that is small for very
large $\left|U_{\rm eff}\right|$, $\delta \varepsilon^{\star}_d
\sim \exp\left[- \left(\pi \left|U_{\rm eff}\right|^2/4 \Gamma
\omega_0 \right)\right]$. \cite{Cornaglia2005a}  In the same limit
the field-dependence of $\delta \varepsilon^{\star}_d$ can also be
evaluated analytically (see Eqs.~(\ref{eq:TAK}) and
(\ref{eq:Jakm}) below). We find
\begin{equation}
\frac{\delta \varepsilon^{\star}_d(B)}{\delta
\varepsilon^{\star}_d(0)} \sim \exp\left[ \pi B^2/\left(16 \Gamma
\omega_0 \right)\right]. \label{eq:deltavB}
\end{equation}
The width of the conductance peak thus increases exponentially
with $B^2$ in the strong coupling limit.

\section{Numerical results} \label{sec:numerical}

In this section we present numerical results for the spin resolved
spectral density $\rho_{d\sigma}(\omega)\!=\!- \pi^{-1}{\textrm
{Im}}\;{\mathcal G}_{d\sigma}(\omega)$ and conductance obtained using
the Numerical Renormalization Group method
(NRG) \cite{Wilson1975,Krishnamurthy1980} modified to include {\it {\it e--ph}}
coupling~\cite{Hewson_2002} and to calculate accurately the spectral density
$\rho_{d\sigma}(\omega)$ using the method of
Ref.~[\onlinecite{Hofstetter2000}].

\begin{figure}[tbp]
\includegraphics[width=8.5cm,clip=true]{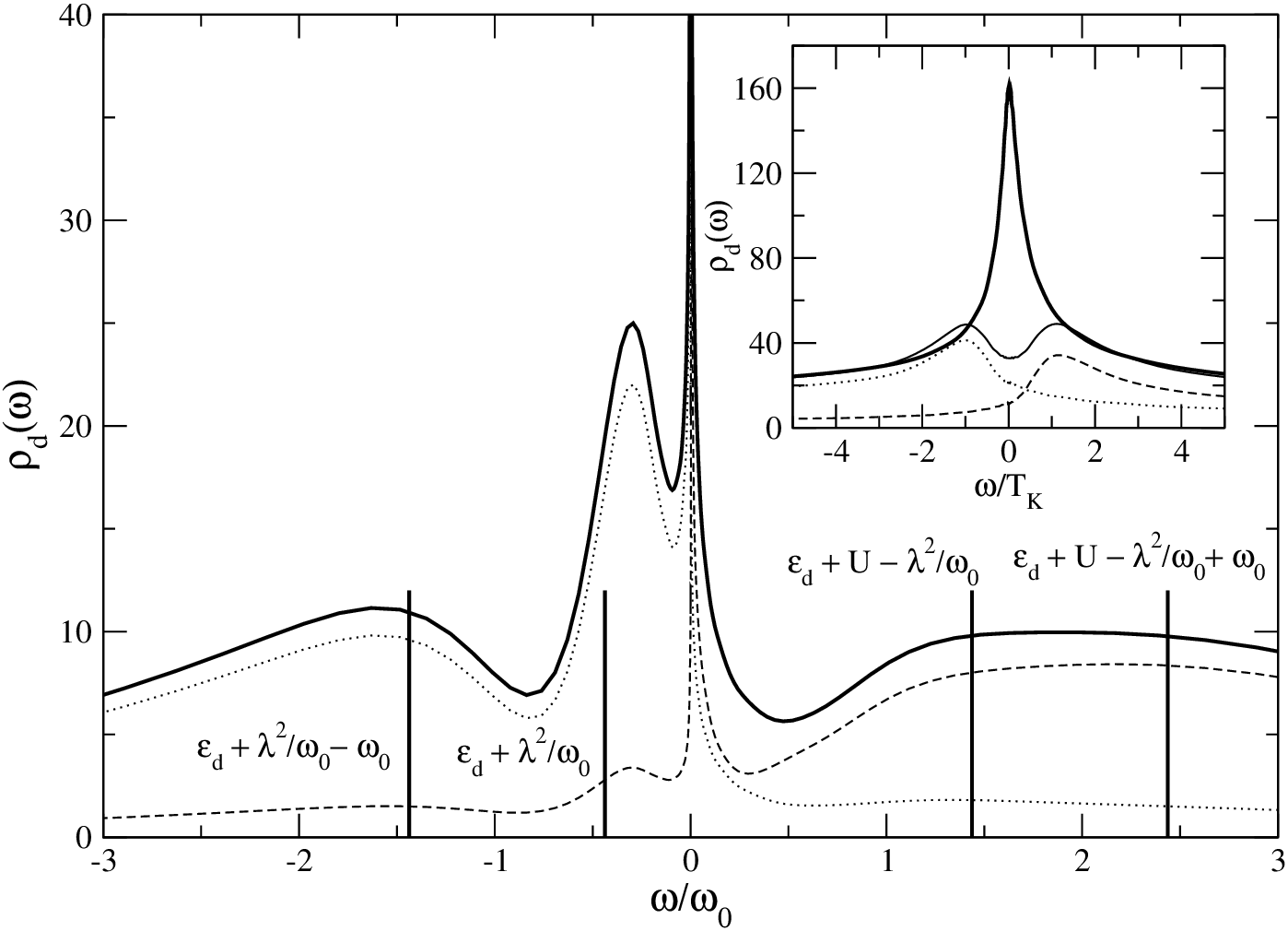}
\caption{Total and spin-resolved spectral densities at $T=0$ for
$U_{\rm eff}\!>\!0$. Parameters values are
$\varepsilon_d\!=\!-0.04$, $\Gamma\!=\!0.00393$, $U\!=\!0.1$,
$\omega_0\!=\!0.02$, and $U_{\rm eff}\!=\!0.0375$. $T_{\rm
K}\!=\!8\times 10^{-5}$. Full line: total spectral density for
$B=0$. Dotted and dashed lines: spin down and spin up components,
respectively, for $B\!\sim\!T_{\rm K}$. Inset: structure of the
Kondo peak. Full thick and thin lines: total spectral density for
$B\!=\!0$ and $B\!\sim\!T_{\rm K}$, respectively. Dotted and
dashed lines: majority and minority spin components of the
spectral density, respectively.} \label{fig:fig1}
\end{figure}
Figure \ref{fig:fig1} represents the total spectral density in
zero field (full line) and its spin resolved components  (dotted
and dashed lines) in the presence of a small magnetic field
$B\!\sim\!T_{\rm K}$ for a case with $U_{\rm eff} > 0$.
Parameters are $\varepsilon_d\!=\!-0.04$, $\Gamma\!=\!0.00393$,
 $U\!=\!0.1$, $U_{\rm eff}\!=\!0.0375$ and $\omega_0\!=\!0.02$.
The system is off the electron-hole symmetric point and
$\delta\varepsilon_d > 0$, $n_d < 1$. The density of states is
thus asymmetric about the Fermi level,
$\rho_d(\omega)\!\ne\!\rho_d(-\omega)$.
\begin{figure}[tbp]
\includegraphics[width=8.5cm,clip=true]{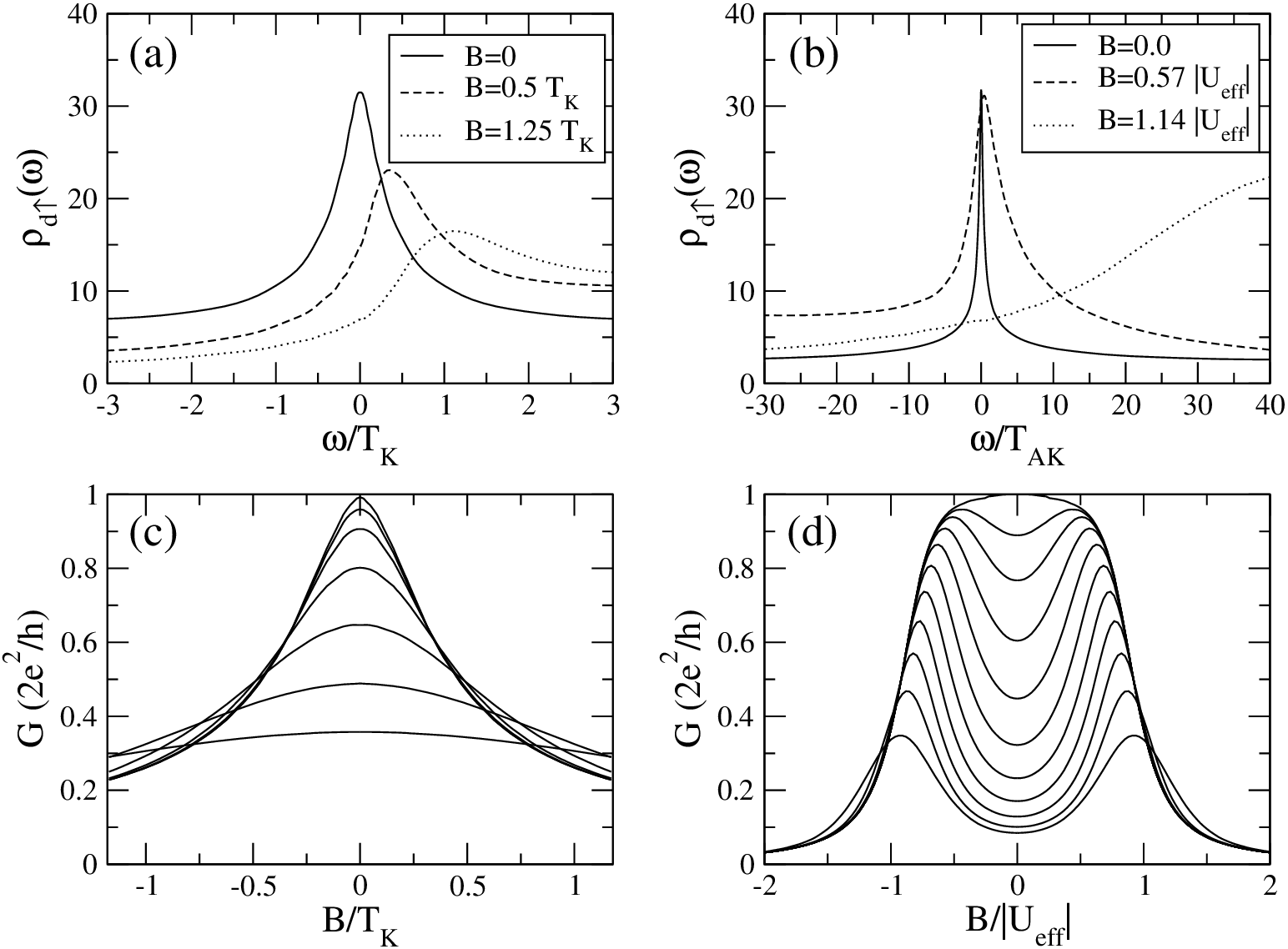}
\caption{Zero temperature spectral density in the region of the
Kondo peak and conductance as a function of temperature and
magnetic field at the electron-hole symmetric point. In all cases
$\varepsilon_d\!=\!-0.05$, $\Gamma\!=\!0.01$, and
$\omega_0\!=\!0.05$. (a): Zero temperature spectral density for
$U_{\rm eff}\!=\!0.051$ and three values of the magnetic field.
$T_{\rm K}\!=\!0.004$. (b) Zero temperature spectral density for
$U_{\rm eff}\!=\!-0.044$ and three values of the magnetic field.
$T_{\rm {AK}}\!=\!1.1\times 10^{-4}$. (c) Conductance as a
function of the magnetic field and temperature for $U_{\rm
eff}=0.051$. Temperatures are $T=0$ (top curve) and $T=0.006
\times 2^{-N}, N=0,1,2,\ldots,5$ (from bottom to top). (c)
Conductance as a function of the magnetic field and temperature
for $U_{\rm eff}=-0.044$. Temperatures are as in (b) with
$N=0,1,2,\ldots,9$.} \label{fig:fig2}
\end{figure}
The total spectral function is represented by the full line. We
observe the presence a very narrow Kondo peak near $\omega\!=\!0$
and additional structure on either side of this feature. The Kondo
temperature, determined from the full-width at half maximum of the
central feature, is $T_{\rm K}\!\sim\!8\times 10^{-5}$ for this
set of parameters. The structure on either side of the Kondo peak
are phonon side bands as can be seen by analyzing the pole
structure of ${\cal G}^{\rm at}_{d\sigma}$, the retarded Green
function in the atomic limit ({\it i.e.}, at $\Gamma\!=\!0$).
\cite{Hewson_2002}

The vertical lines in Fig.~\ref{fig:fig1} represent the positions
of the poles of ${\cal G}^{\rm at}_{d\sigma}$ given in Eq.~(15) of
Reference \onlinecite{Hewson_2002}. It can be seen that they are
in fairly good agreement with the positions of the peaks observed
away from the Kondo feature in Fig.~\ref{fig:fig1}.

The presence of a field $B\!\sim\!T_{\rm K}$ has drastic
effects on the spin resolved spectral spectral function. Due to
the high magnetic susceptibility of the molecule ($\propto
1/T_{\rm K}$) the latter acquires  a large magnetization. Most
of the electron density is composed of majority spins (down spins)
and the population of the minority spins (up spins) is small. The
spectral function of the majority spins (represented by dotted
lines) is thus large for $\omega\!<\!0$ whereas that of the
minority spins (represented by dashed lines) is large for
$\omega\!>\!0$.

The inset to the figure shows in detail the effect of the magnetic
field on the Kondo peak. The thick and thin full lines represent
the total spectral function in the central region in zero field
and  for $B\!\sim\!T_{\rm K}$, respectively. It can be seen that
the Kondo peak splits and its amplitude is reduced anticipating
the suppression of the Kondo effect at higher fields. The dotted
and dashed lines represent the spin down and spin up components of
$\rho_d(\omega)$, respectively. It is seen that position of the
two Kondo peaks shift in opposite directions by an amount $\sim
\pm B$.  \cite{Costi2001}

Figure \ref{fig:fig2} displays the spin resolved spectral density
at $T\!=\!0$ for three values of the magnetic field (a,b), and the
conductance as a function of  magnetic field and temperature (c,d)
at the electron-hole symmetric point. The parameters are
$\Gamma\!=\!0.01$, $U\!=\!0.1$, and $\omega_0\!=\!0.05$. We only
show the minority-spin spectral density $\rho_{d\uparrow}$ since
electron-hole symmetry and spin reversal invariance lead to the
relationship
$\rho_{d\downarrow}(\omega)=\rho_{d\uparrow}(-\omega)$.

It can be seen that the Kondo peak behaves differently in a
magnetic field in the cases shown in Fig.~{\ref{fig:fig2}}(a)
($U_{\rm eff}\!=\!0.051\!>0$) and Fig.~{\ref{fig:fig2}}(b)
($U_{\rm eff}\!=\!-0.044\!<\!0$). Whereas in the former case the
Kondo peak shifts and its hight decreases upon applying a small
magnetic field, in the latter case, for fields $B\!\lesssim
\left|U_{\rm eff}\right|$, its position and height remain
unchanged. Note that the shape of the peak is asymmetric and its
width increases sharply upon applying the field. The spectral
asymmetry is due to the decrease of the density of minority spins
in the presence of the field as discussed above. The
field-dependence of the width of the Kondo peak will be discussed
below.

For even larger fields, $B\!>\left|U_{\rm eff}\right|$,
there is a level crossing and the charge doublet is no longer the
 ground state of the isolated molecule. The charge Kondo effect can not
take place and the associated peak in the density of states
disappears.
\begin{figure}[tbp]
\includegraphics[width=8.5cm,clip=true]{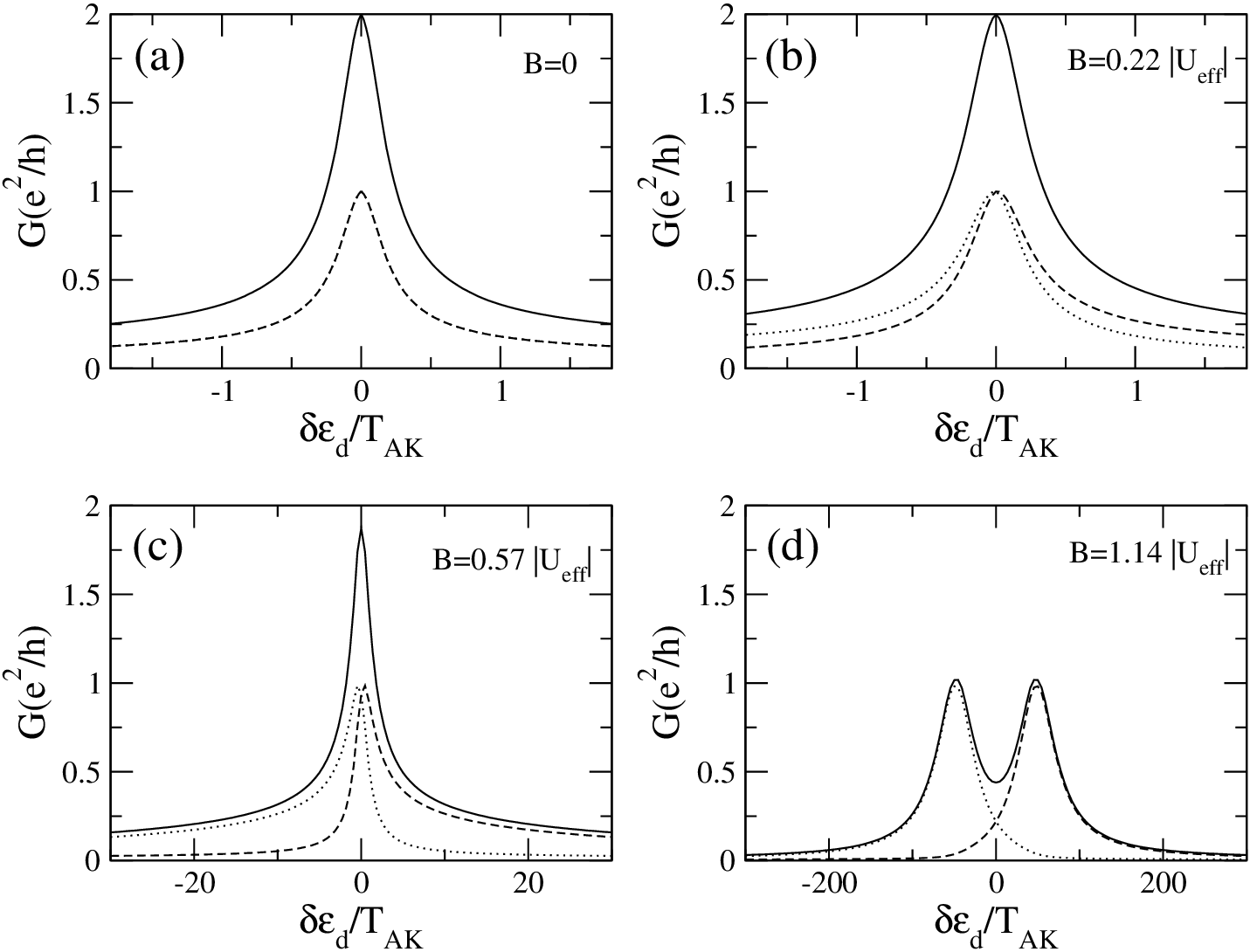}
\caption{Zero temperature conductance $G$ versus gate voltage
 $\varepsilon_d$ for different values of the magnetic field B in the strong {\it
e--ph} coupling regime (a-d). Solid lines: total conductance.
Dashed and dotted lines: majority (down) and minority (up) spins, respectively.
Other parameters as in Fig \ref{fig:fig2}(b).} \label{fig:fig3}
\end{figure}

The dependence of the conductance on field and temperature for the
cases of positive and negative $U_{\rm eff}$ is shown in
Figs.~\ref{fig:fig2}(c) and \ref{fig:fig2}(d), respectively.

For positive $U_{\rm eff}$, the field-induced suppression of
spectral weight at the Fermi level seen in Fig.~\ref{fig:fig2}(a)
translates into a suppression of the zero temperature conductance
as can be seen in Fig \ref{fig:fig2}(c). As a function of field
this quantity exhibits a peak of width $\sim \;T_{\rm K}$, as
expected from Eq.~(\ref{g-de-B}). With increasing temperature the
conductance decreases further.

The case of  negative $U_{\rm eff}$ is shown in Fig.
\ref{fig:fig2}(d). At temperatures much lower than the zero-field
value of $T_{\rm {AK}}$ we observe a wide plateau of width $\sim 2
|U_{\rm eff}|$. The region of the plateau is that in which charge
Kondo effect exists as discussed above and this leads to a high
conductance. With increasing temperature the conductance first
decreases in the central region where $T_{\rm {AK}}$ is the lowest
(see below).

At high temperatures, two peaks are seen at $B\sim\pm |U_{\rm
eff}|$. These peaks are associated to the crossing of the levels
of the isolated molecule [cf. Eq.~(\ref{eq:enes})]: for
$\left|B\right|\!>\!\left|U_{\rm eff}\right|$ the ground state of
the molecule is either $\left|\downarrow,0\right.\rangle$ or
$\left|\uparrow,0\right.\rangle$, depending on the sign of the
field. For $\left|B\right|\!<\!\left|U_{\rm eff}\right|$, instead,
the ground state of the molecule is spinless. These peaks are thus
the analogs of the familiar Coulomb blockade peaks that appear for
positive $U_{\rm eff}$ upon varying the charge. For negative
$U_{\rm eff}$ and in the electron-hole symmetric situation that we
are discussing the charge is fixed, however, and it is the spin of
the molecule that changes at the peaks by an amount $\Delta S_z =
1/2$ when the magnetic field varies.

We now discuss the spin resolved conductance at $T\!=\!0$ as a
function of gate voltage ({\it i.e.}, of $\varepsilon_d$) for the
case $U_{\rm eff}\!<\!0$ and several values of the magnetic field.
Figure \ref{fig:fig3}(a) shows the conductance at zero field. When
the molecular level departs from the symmetric position, the
molecule charge--polarizes and the conductance decreases according
to Eq. (\ref{g-de--e}). As a consequence, the conductance exhibits
a narrow peak of width $\sim \;T_{{\rm {AK}}}$.

When a small magnetic field is applied, $T_{{\rm {AK}}}$ increases
as mentioned above and the peak in the conductance broadens
accordingly [cf. Fig. \ref{fig:fig3}(b)]. There is also a spin
polarization of the current that changes sign at the symmetric
point. This can be understood as follows: for negative but small
$\delta\varepsilon_d$ the charge of the molecule increases and the
current is dominated by charge fluctuations between the states
$\left|2,0\right\rangle$ and $\left|\downarrow,0\right.\rangle$.
This leads to a predominantly spin-up current. For positive but
small $\delta\varepsilon_d$ the charge in the molecule decreases
and the relevant fluctuations are now those between
$\left|0,0\right.\rangle$ and $\left|\downarrow,0\right.\rangle$
and the spin-polarization of the current is reversed.

The width of the conductance peak further increases with the
magnetic field but the peak height remains remains nearly
unchanged as seen in Fig.~\ref{fig:fig3}(c). Notice the strong
spin-polarization of the current reflected in the strong asymmetry
of the peaks in the spin resolved conductance. In this regime the
device acts as a very effective spin filter and the spin
polarization can be reversed by a small change in the gate
voltage.
\begin{figure}[tbp]
\includegraphics[width=8.5cm,clip=true]{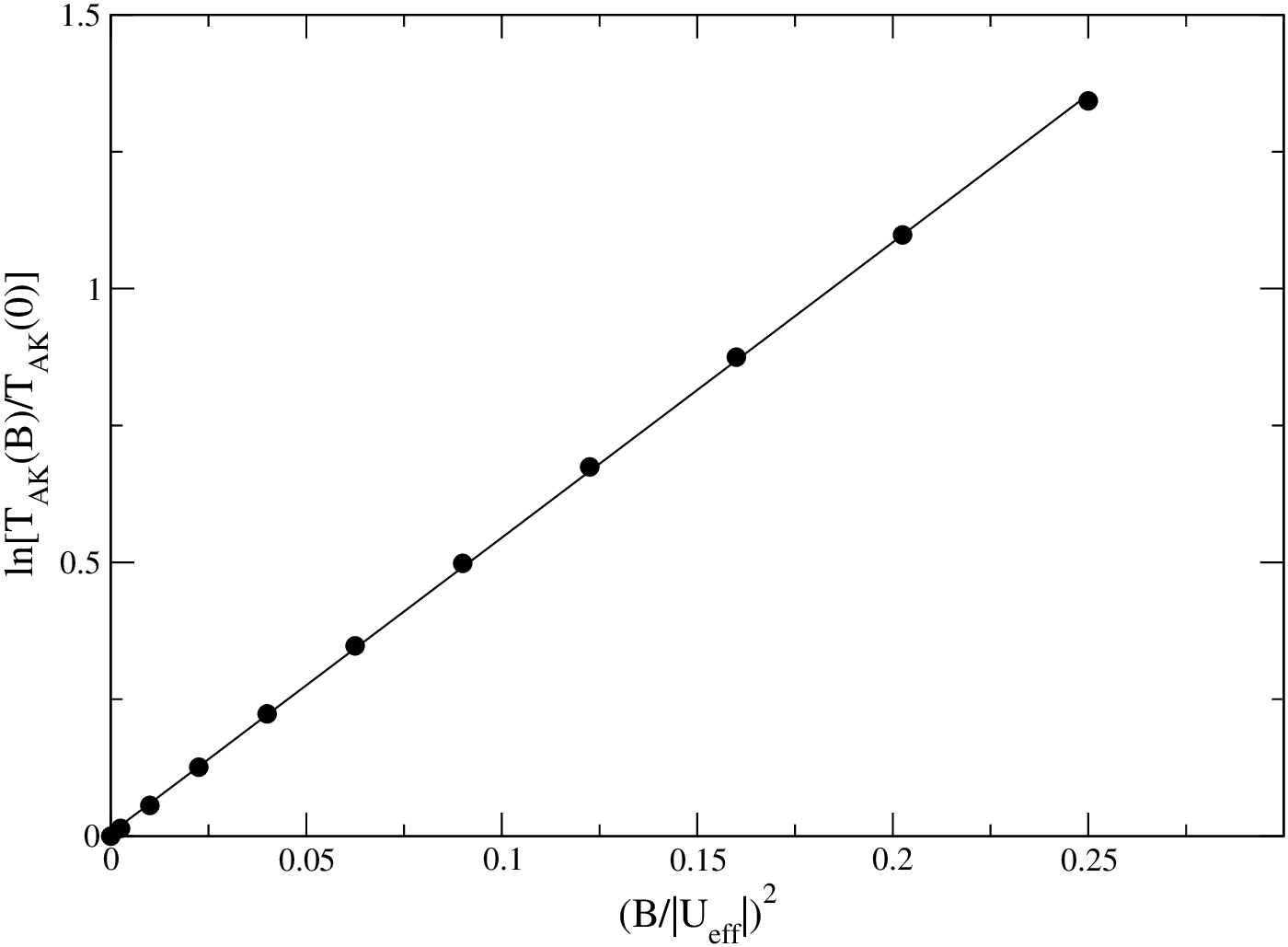}
\caption{Magnetic field dependence of $T_{{\rm {AK}}}$ computed
from Eqs.~(\ref{eq:Jakm}) and (\ref{eq:TAK}) for the same values
of the parameters as in Fig.~\ref{fig:fig3}. The thin line is a linear fit.} \label{fig:fig4}
\end{figure}
For magnetic fields larger than the spin gap $B\gtrsim
|U_{\rm eff}|$, the ground state of the isolated molecule
changes as mentioned above and the total conductance shows a
two-peak structure. The heights of these are reduced by half as
shown in Fig.~\ref{fig:fig3}(d). These peaks are located at
$\delta\varepsilon_d \sim \pm\;\left(B -|U_{\rm eff}|\right)/2$
where there is a level crossing of the molecular states. The
conductance peaks are almost fully spin-polarized in this limit.

We conclude by discussing in detail the field-dependence of the
charge Kondo temperature $T_{{\rm {AK}}}$. As shown in Reference
\onlinecite{Cornaglia2005a}, the low-energy effective model for
the charge fluctuations in the strong {\it e--ph} phonon coupling
case is the anisotropic Kondo model with couplings $J_{\parallel}$
and $J_{\perp}$ and Kondo temperature~\cite{Costi1996, Costi1998}
\begin{equation}
T_{{\rm{AK}}} \sim \left({J_\perp/J_\parallel}\right) ^{1\over
J_\parallel \rho_0}. \label{eq:TAK}
\end{equation}

In the present case, in the presence of a magnetic field, the
coupling constants are \cite{Cornaglia2005a}
\begin{widetext}
\begin{equation}
J_{\parallel} =  \frac{J_0}{2}\!\sum_{m,\sigma\!=\!\pm 1/2}
{e^{-\left(\lambda/\omega_0\right)^2}
\left(\lambda/\omega_0\right)^{2 m}/m! \over
 {2\lambda^2/\omega_0 U} -1 -2 \sigma B/U + {2 m \omega_0/U}}\;,\;\;\; J_{\perp} =
\frac{J_0}{2}\!\sum_{m,\sigma\!=\!\pm 1/2} {(-1)^m
e^{-\left(\lambda/\omega_0\right)^2}
\left(\lambda/\omega_0\right)^{2 m}\!/m! \over
 {2\lambda^2/\omega_0 U} -1 - 2 \sigma B/U + {2 m \omega_0/U}},
\label{eq:Jakm}
\end{equation}
\end{widetext}
where $J_0\!=\!8\;\Gamma/(\pi U \rho_0)$.

For very large {\it e--ph} coupling the expressions above can be
evaluated analytically leading to the result quoted in
Eq.~(\ref{eq:deltavB}) in Sec. \ref{sec:model}. In the numerical
calculations reported here we used moderate values of $\lambda$
and the expressions for the couplings in Eq.~(\ref{eq:Jakm}) must
be evaluated numerically.

Figure \ref{fig:fig4} represents the field dependence of
$T_{{\rm{AK}}}$ for the same parameters as in Fig.~\ref{fig:fig3}.
It can be seen that in this regime $T_{{\rm{AK}}}$ still increases
exponentially with $B$.

\section{Conclusions and summary}\label{sec:conclusions}
We analyzed the equilibrium transport and spectral properties of a
model molecular transistor when the spin symmetry is broken by applying
an external magnetic field. For weak {\it e--ph} coupling, in zero
field, the conductance shows the well known Coulomb blockade peaks
at high temperature and the enhancement of the conductance due to
the Kondo effect at low temperatures. An external
 magnetic field $B\gtrsim T_{\rm K}$ suppresses the conductance as
 a result of the splitting of the Kondo peak.
 The properties of the conductance at zero bias
 are qualitatively the same as those found in the case in which there is no {\it e-ph}
 coupling but with a set of phonon renormalized parameters.
 The spectral density, however, presents a
set of additional peaks separated by the phonon energy $\omega_0$,
that can be understood analyzing the level structure of the
isolated molecule. The properties of the spectral density are
experimentally accessible in transport experiments at finite but
small source-drain voltages or in scanning tunneling microscopy
(STM) experiments for molecules on surfaces.

In the strong {\it e--ph} coupling regime the behavior of the
system is qualitatively different. The odd Coulomb blockade
valleys are completely suppressed and the low temperature physics
is dominated by charge fluctuations instead of spin fluctuations.
In a small range of gate voltages the charge Kondo effect takes
place and results in an enhancement of the conductance at low
temperatures. In contrast with the previous case, the Kondo peak
is not split by a magnetic field. Instead, for a broad range of
magnetic fields, it broadens as the latter increases. For large
enough fields or temperatures a {\it spin blockade} regime is
observed.

The magnetic field dependence of the conductance of a molecular
transistor with strong {\it e--ph} coupling resembles in many ways
the gate voltage dependence of the conductance of a device with
weak {\it e--ph} coupling in zero field. The two cases are related
to each other through the transformation $\langle \hat{S}^z_{d}
\rangle$ $\to$ $\langle \hat{n}_d \rangle$ and $B \to V_g$. This
transformation is not an exact one because the symmetries in the
spin and charge sectors of the Hilbert space are different but it
gives a good qualitative physical picture.

\bibliography{refs}
\end{document}